\documentstyle[pra,aps]{revtex}
\begin{document}
\draft
%\twocolumn
%\voffset=-0.75in
\title{Limits for compression of quantum information carried by ensembles of 
mixed  states}

\author{Micha\l{} Horodecki \cite{poczta1}}

\address{Institute of Theoretical Physics and Astrophysics\\
 University of Gda\'nsk, 80--952 Gda\'nsk, Poland}

\def\hcal{{\cal H}}
\def\ecal{{\cal E}}
\def\ccal{{\cal C}}
\def\qcal{{\cal Q}}
\def\pcal{{\cal P}}

\def\dim{{\rm dim}}
\maketitle

\begin{abstract}
We consider the problem of compression of  the quantum information carried 
by ensemble  of mixed states. We prove that for arbitrary coding schemes 
the least number of qubits needed to convey the signal states
asymptotically faithfully is bounded from below by the Holevo function 
$S(\varrho)-\sum_ip_iS(\varrho_i)$.  We also show that a compression 
protocol can be composed with another one, provided that the latter 
offers {\it perfect} transmission.  Such a compound protocol is applied
to  the case of  binary source.  It is conjectured to reach the obtained
bound.  Finally, we point out that in the case of mixed signal
states there could be a difference between the maximal compression 
rates at the coding schemes which are ``blind'' to the signal and the 
ones which assume the knowledge about the identities of the signal states.
\end{abstract}


\newpage

\section{Compression of quantum information}
\label{wstep}
One of the important problems of information theory is the compression of
information. A general limit for compression rate of classical information is
placed by the so called noiseless coding theorem \cite{Shannon}. Suppose that
a source generates a message $i$ with probability $p_i$ and allow to cumulate
the subsequent messages into long sequences  and then represent them (encode)
as sequences of bits as economically as it is possible (economically means
here that we want to use the least possible average number of bits per
message). The task of the receiver (Bob) is to convert (decode) the binary
sequences into the original sequences of messages. 
Here we do not require perfect transmission but only asymptotically faithful
transmission. This means that Bob may be unable to recover correctly each
sequence, but the probability of error tends to zero if the length 
of input blocks tends to infinity.

Now the noiseless coding theorem \cite{Shannon} says that the
necessary and sufficient number of bits per message needed for assymptotically
faithful transmission is equal to the Shannon entropy
$S=-\sum_ip_i\log p_i$  (in this paper we use base-2 logarithms)
of the probability
distribution characterizing the source. Then this quantity says in fact
how much information
per message is actually produced by the source. Indeed, one can imagine, that
after the most economical compression procedure, each piece of the
compressed signal is now equally essential as all redundancy was removed.
Then the size
of the maximally compressed signal can be interpreted as the quantity of 
information
contained in the input (uncompressed) one.

Let us now turn to the problem of compression of quantum information which
was first considered by Schumacher \cite{Schumacher95}.
The messages are here replaced by quantum states $\varrho_i$ and the bits
by qubits i.e. two-level systems. The probability of error is
generalized to quantum case by means of the chosen measures of
fidelity or distortion
\cite{Schumacher95,Jozsa,Lo} between two quantum states.
Thus we will ask about the least number
of the two level systems needed to carry the  information
assymptotically faithfully to Bob i.e. so that the average distortion
between the input and output states will
tend to zero (or the average fidelity will tend  to one) in the limit of
input signal block  of infinite length.

Before we review the results obtained so far, let us mention the fundamental
difference between the quantum and classical case  due to the 
no-cloning theorem for quantum states \cite{Wooters}. 
It was shown that the theorem is  equivalent to the impossibility of
measuring of the  state parameters of single quantum system \cite{Yuen}.
Then we can imagine two scenarios which, according to the above restriction for 
quantum information processing, could in principle produce different results
\cite{Barnum}.
Within the first scenario, we assume that Alice does not know the 
identities of the
particular states produced by the source. Then, in accordance with
the no-cloning theorem,
Alice has  no means to get this knowledge. Thus   the most general Alice's
coding protocol amounts to performing a quantum operation (trace preserving
completely positive map - see Appendix) \cite{Kraus} which depends only of
the known characteristics of the
source i.e. the form of the generated ensemble $\{p_i,\varrho_i\}$.
We will call it {\it blind}  coding. However if 
we allow Alice to know each of the produced states, we deal with the
second scenario ({\it arbitrary } or {\it non-blind} coding), where Alice's
coding amounts to replacing the sequences of
signal states by completely  arbitrary new states. It seems that in some 
cases it  will produce more efficient compression than it is possible 
within the previous scenario.

Let us now review the results obtained so far in the domain of compression of
quantum information. For the ensemble of pure states, Schumacher
showed \cite{Schumacher95} that, by means of blind coding,
it is possible to reduce
the needed number of qubits to  the value  of the von Neumann  entropy of
the total density matrix of ensemble  $\varrho=\sum_ip_i\varrho_i$
(in short -  the von Neumann entropy of the ensemble).
The proposed coding-decoding protocol  was then simplified by
Jozsa and Schumacher \cite{JS} (we will refer to it as to SJ protocol).
To obtain the converse statement saying that this quantity is also necessary for
faithful recovery of the signal, Barnum et. al \cite{Barnum}
considered arbitrary coding scheme. It turned out that even in this case
it is impossible to better compress the data, so that the obtained lower bound
applied also for the first scenario.
Thus for the ensemble of
pure states the problem of compression has been completely solved: the two
scenarios give the same degree of compression, and the information per
message contained in such an ensemble is equal to the von Neumann entropy
$S(\varrho)$  of
the ensemble. Note that this establishes a precise sense of the
von Neumann entropy within the quantum information theory \cite{thermo}.

Now, the problem of ensemble of {\it mixed} states is still open. The
SJ coding protocol allows to compress such an
ensemble down to the value of its von Neumann entropy \cite{Lo,Jozsa_mix}
(see also \cite{Armenia} in this context), but
one knows that in some cases the more efficient protocols are possible
\cite{Lo,Jozsa_mix}. To illustrate it, let us  consider  the source
producing with certainty some established mixed state. Then the ensemble
has entropy greater than zero but of course it does  not carry any information.
This
implies that the ``information content'' of the ensemble  (for any of the two
scenarios) cannot be, in general,  merely a function of the density matrix of
the ensemble. Instead, it must depend on the particular form of the ensemble.
Moreover, it seems that for ensembles of mixed states the arbitrary coding
could produce more efficient compression than the blind one.
Under the consideration it is  desirable to investigate the problem of
compression of information carried by ensembles of mixed states.
In particular, an important task is to provide
some limits for the compression rates with the two types of coding.

In this paper we provide the lower bound for 
the necessary number of qubits per message needed for faithful transmission
of the quantum information carried by an ensemble of mixed states for
arbitrary coding. 
The bound is equal to the function
$S(\varrho)-\sum_i p_i S(\varrho_i) $ (we will call it Holevo information)
which was shown by Holevo to be an upper bound for accessible
information \cite{Holevo,Ozawa}. In particular it implies that for the
ensembles of states of disjoint supports the two considered
types of coding produce the
same result. 
Further we investigate the problem of composing of the compression protocols. 
We consider a class of non-blind coding protocols,
which involve
composition of two protocols: an ideal one, which amounts to replacing the
input states by the new states which, partially traced, reproduce the former
ones, and the SJ protocol (applied to mixed states).
Finally, we conjecture that if the arbitrary coding schemes are allowed then
the Holevo information is in fact equal to
the minimal number of qubits needed for faithful transmission
and the bound can be reached by means of  the proposed class of protocols.

\section{Compression protocols}
\label{protocols}
Suppose that Alice generates a signal state $\varrho_i^0$ acting on a
Hilbert space ${\cal H}_{\cal Q}$ with probability $p_i^0$. The produced
ensemble $\ecal_0=\{p_i^0,\varrho_i^0\}$ has the density matrix
$\varrho^0=\sum_ip_i^0 \varrho_i^0$. Denote now the product
$\varrho_{i_1}^0\otimes\ldots\otimes \varrho_{i_N}^0$ by $\varrho_i$, where
$i$ now stands for multiindex (to avoid complicated notation we  do not write
the index $N$ explicitly unless necessary).
The corresponding ensemble and state are
denoted by $\ecal$ and $\varrho$ respectively. Now Alice performs a coding
operation over the initial ensemble $\ecal_0$ ascribing to any input state
$\varrho_i$ a new state $\tilde\varrho_i$. The map $\varrho_i \rightarrow
\tilde\varrho_i=\Lambda_A(\varrho_i)$ is supposed to be a quantum operation
for blind coding or an arbitrary map - for non-blind one.
In the latter case
we allow Alice even to know which states are generated by the source, so
that she can prepare separately each of the states $\tilde\varrho_i$ for
each $i$.

The new states $\tilde \varrho_i$
represent the compressed signal which is then  flipped into
the suitable number of qubits determined by the dimension of subspace occupied
by the state $\tilde\varrho$ of the ensemble and sent through the
noiseless channel to Bob.
Now the states
$\tilde\varrho_i$ are to be decoded to become close to the initial states
$\varrho_i$. For this purpose Bob performs some established quantum
operation $\Lambda$ which of course does not depend on $i$. Then the resulting
states are $\varrho'_i=\Lambda_B(\tilde\varrho_i)$ and the whole scheme is the
following
\begin{equation}
\varrho_i \quad
\mathop{\longrightarrow}
\limits_{\Lambda_A}^{{\rm Alice's \ coding}} \quad
\tilde\varrho_i \quad
\mathop{\longrightarrow}
\limits_{I}^{{\rm noiseless \  channel}} \quad
\tilde\varrho_i \quad
\mathop{\longrightarrow}
\limits_{\Lambda_B}^{{\rm Bob's  \ decoding}} \quad
\varrho'_i
\end{equation}
where $\varrho_i$ and $\varrho_i'$ act on the Hilbert space
$\otimes^N\hcal_\qcal$ while $\tilde\varrho_i$ on the channel Hilbert space
$H_{\ccal}$. Without loss of generality, we can assume (as in Ref.
\cite{Barnum}) that $H_{\ccal} =\otimes^N\hcal_\qcal$.
%The sequence of the pairs $(\Lambda_A,\Lambda_B)$
%(recall that the pair is implicitly indexed by N) we will call protocol.
As a measure of distortion characterizing the quality of the transmission
$\varrho_i\rightarrow\varrho_i'$  we choose the metric induced by the
trace norm. The latter is defined as
\begin{equation}
\|A\|={\rm Tr}|A|
\end{equation}
with $|A|=\sqrt{A^\dagger A}$. Thus the trace norm of Hermitian operator
is simply the sum of absolute values of its eigenvalues. Consequently, the
distortion is defined as
\begin{equation}
D(\varrho,\sigma)=\|\varrho-\sigma\|
\end{equation}
An important property of the proposed measure of distortion is the fact that
it does not increase under the quantum operations (see Appendix).
Then the average distortion
$\overline{D}\equiv\sum_ip_i
D(\varrho_i,\varrho'_i)$ will
indicate us the quality of the process of recovery  of quantum information by
Bob after compression by Alice.
Now, for a fixed source, determined by the ensemble $\ecal_0$ one considers
the sequence of coding-decoding pairs $(\Lambda_A,\Lambda_B)$
with the property that
$\lim_{N\rightarrow\infty}\overline{D}=0$
(recall that the pair is implicitly indexed by N). 
Such sequences will be called protocols.

Define now the quantity $R_P$ characterizing the asymptotic degree
of compression of the initial quantum data at a given protocol $P$ by
\begin{equation}
R_P=\lim_{N\rightarrow\infty}{1\over N}\log \dim\tilde \varrho
\end{equation}
Here $\dim\tilde \varrho$ denotes the dimension of the support of the state
$\tilde\varrho$ given by  the number of nonzero eigenvalues. The
quantity $\log\dim\tilde\varrho$ has the interpretation of the number of
qubits needed to carry the state $\tilde\varrho$ undisturbed
($\tilde\varrho$ is to be transferred by a noiseless channel).

Now, given a class $\pcal$ of protocols, we define the quantity
\begin{equation}
I_{\pcal}=\inf_{P\in\pcal}R_P
\end{equation}
which is equal to the least number of qubits per message needed for
asymptotically faithful transmission of the initial signal states from Alice
to Bob within the considered class of protocols (to be strict one needs
$I_{\pcal}+\delta$ qubits per message, where $\delta$ can be chosen arbitrarily
small).  As discussed in sec. \ref{wstep}, we are interested in two
classes of protocols - the ones with blind and arbitrary coding
schemes. Accordingly we will consider two kinds of informations - the {\it
passive information} $I_p=I_{\pcal}$  where $\pcal$ is the class of protocols
with blind coding and the {\it effective
information} $I_e$ with the infimum taken over protocols with
arbitrary coding.
The effective information  represents the amount of information which
seems to be {\it actually} carried by the ensemble while the passive information
$I_p$ 
represents the information which is ``seen'' by the quantum apparatus which
is ``blind'' to the signal. Although the actual information contents of the
ensemble could be in fact lower, the apparatus cannot benefit it, as it cannot
in general read the identities of the signal states without
disturbance. In result the compression rate is
restricted by the value of passive information.
Finally  it is convenient to introduce the {\it information defect}
$I_d=I_p-I_e$. This quantity 
says us how  the ensemble is ``unkind'' to us:
while carrying little information the ensemble requires to be  processed as
if it contained a large amount of information.
 Let us recall	here that for ensemble of pure states the
impossibility of reading of the input states does not decrease the
compression efficiency and the defect is equal to zero in spite
of nonorthogonality of the signal states $\cite{Barnum}$.

\section{The bound for effective information}
\label{twierdzenie}
In this section we will prove the main result of this paper.

{\bf Theorem.-} The Holevo information
$I(\ecal_0)=S(\varrho)-\sum_ip_iS(\varrho_i)$ of the ensemble is the lower
bound for its effective information:
\begin{equation}
I_e(\ecal_0)\geq I(\ecal_0)
\end{equation}

Note that, since by  definition we have $I_e\leq I_p$ then the theorem
provides automatically
the lower bound for passive information $I_p$. Note also that for ensembles of
pure states the Holevo information is simply equal to the entropy of
the ensemble so that the theorem is compatible with the result
of Ref. \cite{Barnum} (up to the measure of the quality of
transmission).

To prove the theorem we need the lemma saying that if the average distortion
between the two ensembles is small, then the difference between
their Holevo informations per message is also small.

{\bf Lemma.-} Let $\sum_ip_i\|\varrho_i-\varrho_i'\|=\epsilon\leq{1\over2}$.
Then the following inequality is valid
\begin{equation}
|I(\ecal)-I(\ecal')|\leq2\left[\epsilon N\log d+\eta(\epsilon)\right],
\label{lemat}
\end{equation}
where $\eta(x)=-x\ln x$ with $\eta(0)=0$, $d=\dim\hcal_\qcal$.

{\bf Proof.-} We will use  the following estimate \cite{Fannes}
\begin{equation}
|S(\varrho)-S(\sigma)|\leq\|\varrho-\sigma\|\log\dim\hcal +\eta
(\|\varrho-\sigma\|)
\end{equation}
which is valid for states $\varrho$, $\sigma$ acting on the Hilbert space
$\hcal$, with $\|\varrho-\sigma\|\leq{1\over2}$.
Basing on the above inequality, we obtain
\begin{eqnarray}
&&|S(\varrho)-S(\varrho')|\leq N\log d\|\varrho-\varrho'\|+
\eta(\|\varrho-\varrho'\|)\nonumber\\
&&\leq N\log d \sum_ip_i\|\varrho_i-\varrho_i'\|+\eta(\sum_ip_i\|\varrho_i-
\varrho_i'\|)=\epsilon N\log d +\eta(\epsilon)
\end{eqnarray}
where we used the fact that the trace norm is  convex and that
the function $\eta$ is increasing on the interval $(0,{1\over2})$.

We also have
\begin{eqnarray}
&\sum_ip_i|S(\varrho_i)-S(\varrho_i')|&\leq \sum_ip_i\left[
N\log
d\|\varrho_i-\varrho_i'\|+\eta(\|\varrho_i-\varrho_i'\|)\right]\nonumber\\
&&\leq\epsilon N \log d +\eta(\epsilon)
\end{eqnarray}
where the concavity of the function $\eta$ was used. Now adding the two above
inequalities we obtain the desired result.

Now we can start to prove the theorem. For this purpose
let us estimate the quantity $\log
\dim\tilde\varrho$. First, it is bounded from below by $I(\tilde \ecal)$. This
follows from the obvious  fact that the von Neumann entropy of a state
is less than or equal to the  logarithm of the dimension of the Hilbert space
the state acts on. Now let us note \cite{Ozawa,Scutaru} that the function
$I(\ecal)$ can be written as the mean relative entropy between the
components $\varrho_i$ of ensemble and the density matrix $\varrho$ 
of the latter
\begin{equation}
I(\ecal)=\sum_ip_i S(\varrho_i|\varrho)
\end{equation}
where the relative entropy \cite{Umegaki} is given by
\begin{equation}
S(\varrho|\sigma)={\rm Tr}(\varrho\log\varrho -\varrho\log \sigma)
\end{equation}
Then we can benefit the Uhlmann monotonicity theorem
\cite{Lindblad} which states that the relative entropy does not
increase under the action of completely positive trace preserving map (quantum
operation).
Thus we obtain the inequality
\begin{equation}
I(\tilde\ecal)\geq I(\ecal')
\label{uhl}
\end{equation}
as the the ensemble $\ecal'$ is produced by Bob's
quantum operation from	 the ensemble $\tilde\ecal$.
Using the inequality (\ref{uhl}) and applying the lemma
we get
\begin{equation}
\log\dim\tilde\varrho\geq I(\ecal)-2\bigl[\overline{D} N\log\dim
\hcal_{\qcal}+\eta(\overline{D})\bigr].
\end{equation}
%where $\epsilon=\sum_ip_iD(\varrho_i,\varrho'_i)$
Noting that $I(\ecal)=N I(\ecal_0)$, dividing both sides of the obtained
inequality by $N$ and taking the limit $N\rightarrow \infty$ we
obtain the desired result.

Let us now summarize the idea of the  proof.
First, the number of needed qubits per message is
bounded from below by $I(\tilde\ecal)/N$. Now
Bob obtains the final ensemble $\ecal'$ from the ensemble $\tilde\ecal$
by means of quantum operation which by Uhlmann theorem can only
decrease the Holevo information per message.
Hence we have $I(\tilde\ecal)/N\geq I(\ecal')/N$.
But from the lemma it follows that the initial and final
ensembles have asymptotically equal Holevo information per message
$I(\ecal)/N\approx I(\ecal')/N$ hence we obtain
$I(\tilde\ecal)/N\geq I(\ecal)/N$ in the limit of large $N$.
Note here that if the bound is to be reached, then the
asymptotic mean entropy of the ensemble $\tilde\ecal$ per message must vanish.
This follows from the fact that the estimate of the $\log\dim\tilde\varrho$
by the Holevo information is not too rough only if the latter amounts
to the von Neumann entropy.

Finally note that for the case
of blind coding the Holevo information per message must be equal for all
three ensembles $\ecal$, $\tilde\ecal$, $\ecal'$. In other words we can say
that the Holevo information is	{\it invariant} under the asymptotically
reversible operations. The same {\it cannot} be stated for von Neumann entropy.
Indeed, otherwise we would not be able to compress the signal more than
indicated
by the von Neumann entropy. However we know that it is possible e.g. for
a particular ensemble considered in Ref. \cite{Lo} which consists of
states of disjoint
support. Here the signal states then can be measured and  replaced by
pure ones. Then the entropy of the ensemble decreases to the value of its
Holevo information. The reversal is done again by measuring the pure states
and replacing them by the initial, possibly mixed ones.
Applying  the theorem we find that the passive and effective information
are equal and take the value of the Holevo information of the ensemble.
Then the information defect vanishes  not only for ensemble of pure states
but also for ensemble of  mixed states with disjoint supports.

%The clue is that
%Alice  uses here {\it non-bistochastic} operation which may decrease the
%von Neumann entropy and which, for a given ensemble, is
%still reversible. Recall that  bistochastic quantum operation \cite{Alicki}
%is the one which preserves the maximally mixed state i.e. the normalized
%identity. Such an  operation can only increase the entropy of the state (this
%follows immediately from the Uhlmann theorem). Therefore only the
%non-bistochastic operation can help us for obtain better compression of mixed
%states. It should be noted here, that the Schumacher-Jozsa coding for each finite
%$N$ is also a non-bistochastic operation. However it becomes asymptotically
%bistochastic i.e. the non-conservation of the maximally mixed state becomes
%negligibly small for large blocks. Thus to be strict we must say about
%asymptotic bistochasticity.

\section{Composing protocols}
\label{puryfikacje}

From the discussion of the previous section it follows that the
entropy of the density matrix of the ``intermediate'' ensemble $\tilde\ecal$
should be as low as possible. In this section we will present a particular
class of non-blind protocols, which aim at decreasing the entropy.
Namely, Alice can replace the input states $\varrho_i$ with such new ones
$\tilde\varrho_i$ acting on larger Hilbert space
$\hcal=\left(\otimes^N\hcal_\qcal\right)\otimes \hcal'$ that
${\rm Tr}_{\hcal'}\tilde\varrho_i=\varrho_i$.
Then the Bob's decoding amounts to performing
partial trace, i.e. discarding the systems described by the Hilbert space
$\hcal'$. Then the states $\tilde\varrho_i$  can produce the
density matrix $\tilde\varrho$ of lower entropy than the initial one.
Clearly, the above scheme provides perfect transmission. However
the matrix $\tilde\varrho$, although of perhaps small entropy,
will usually occupy larger Hilbert space than the source space. To avoid it 
one could compose the present (ideal) protocol with the SJ protocol.
Then the overall scheme is the following
\begin{equation}
\varrho_{i_1}\otimes \ldots\otimes \varrho_{i_k}
\mathop{\longrightarrow}\limits^{{\rm Alice's \atop coding}}
\tilde\varrho_{i_1}\otimes \ldots\otimes \tilde\varrho_{i_k}
\mathop{\longrightarrow}\limits^{{\rm SJ\atop protocol }}
\tilde\varrho_{i_1 \ldots i_k}
\mathop{\longrightarrow}\limits^{{\rm Bob's\atop partial \ trace }}
\varrho_{i_1 \ldots i_k}'
\end{equation}
Here $i_j$'s are multiindices of length $N$;
$\varrho_{i_1}\otimes \ldots\otimes \varrho_{i_k}$ and
$\varrho_{i_1 \ldots i_k}'$ act on the Hilbert space
$\otimes^k\left(\otimes^N \hcal\right)$ while
$\tilde\varrho_{i_1}\otimes \ldots\otimes \tilde\varrho_{i_k}$ and
$\tilde\varrho_{i_1 \ldots i_k}$ act on
$\otimes^k\left(\left(\otimes^N \hcal\right)\otimes\hcal'\right)$.
The latter two states can be obtained from the former ones by tracing over
the space $\otimes^k \hcal'$. As the  used distortion measure
does not increase under the partial trace
operation (see Appendix), the average distortion produced by the
composed protocol is less than or equal to the one within the ``intermediate''
SJ protocol.
The latter distortion tends to zero if $N$ is kept fixed and
$k$ tends to infinity  (of course $N$, although fixed, can be
chosen arbitrarily large).
Then composing the two protocols we have obtained again a compression protocol.
The result can be immediately generalized as follows.
{\it Any protocol
providing perfect transmission can be composed with some other protocol,
so that the full one is again a protocol, i.e. offers asymptotically faithful
transmission.}

Turning back to the considered case, we see that since
the SJ protocol compresses the signal down
to the value of entropy of the source ensemble per message
\cite{Lo,Jozsa_mix}, the following inequality holds 
\begin{equation}
I_p\leq\lim_{N\rightarrow\infty} {1\over N}S(\tilde\varrho),
\end{equation}
where the infimum is taken over the states $\tilde\varrho$ of ensembles, which
partially traced produce the input ensemble $\ecal$.

Let us illustrate the above result  by means of
an example of binary source, i.e. the one which generates two kinds of
messages $\varrho^0_1$ and $\varrho^0_2$ with probabilities $p^0_1$ and
$p^0_2$ respectively (for convenience we will further
omit the indices $0$). Suppose that 
Alice replaces the single signal states by their
purifications $P_i=|\psi_i\rangle\langle\psi_i|$ acting on
the Hilbert space $\hcal_\qcal\otimes\hcal'$ \cite{puri}.
 As the source produces only two kinds of states, the entropy
of the ensemble of purifications can be calculated explicitly
\begin{equation}
S(\tilde\varrho)=H\left[{1\over2}\left(1+\sqrt{(p_1-p_2)^2+4p_1p_2
|\langle\psi_1|\psi_2\rangle|^2}\right)\right],
\end{equation}
where $H(x)=-x\log x - (1-x)\log(1-x)$ is the binary entropy function.
The minimal entropy is obtained if the overlap of $\psi_1$ and $\psi_2$ is the
largest. The conditional supremum of the overlaps
of purifications of the two states $\varrho_1$ and $\varrho_2$
is given by the fidelity of the states  \cite{Uhlmann2,Jozsa}
\begin{equation}
\max|\langle\psi_1|\psi_2\rangle|^2\equiv F(\varrho_1,\varrho_2)=
\left({\rm Tr}\sqrt{\sqrt{\varrho_1}\varrho_2\sqrt{\varrho_1}}\right)^2,
\end{equation}
so that we obtain
\begin{equation}
I_e\leq S_{\min}(\tilde\varrho)=H\left[{1\over2}\left(1+
\sqrt{(p_1-p_2)^2+4p_1p_2
F(\varrho_1,\varrho_2)}\right)\right].
\end{equation}
Now if $\varrho_1$ and $\varrho_2$ have disjoint supports, then
$F(\varrho_1,\varrho_2)=0 $ and $S_{\min}$ is equal to the Holevo
information of the ensemble, which is compatible with discussion in
sec. \ref{twierdzenie}.
(and discussion in Ref. \cite{Lo}).
Note that the presented protocol is performed {\it separately} on the single
messages. It seems reasonable to conjecture that  if the protocol was applied
to the blocks of messages then one could reach the bound of Holevo information
for general ensembles.
In other words it is very probable  that in fact
$I_p=I(\ecal_0)=\lim_{N\rightarrow\infty} {1\over N}\min_{\tilde
\varrho}S(\tilde\varrho)$.
However, it is difficult to calculate the minimal asymptotic
entropy per message even for the case of binary source.

\section{Conclusion}

In conclusion, we have considered the problem of  compression of quantum
information carried by an ensemble of mixed states. We have proved that the  
minimal number of qubits per message needed for asymptotically faithful
transmission is greater than the Holevo information of the initial ensemble.
We have also showed that any protocol providing perfect transmission can
by successfully composed with another protocol.
We proposed a non-blind protocol involving composition of a
perfect protocol with the Schumacher-Jozsa one. The first stage bases
on replacing the signal states by the new states which, partially traced,
reproduce the initial ones.
The proposed scheme, if applied to blocks of messages,
is conjectured to reach the bound.
Then the Holevo information would acquire the physical sense within the
quantum information theory, being a proper generalization of von Neumann
entropy to the case of ensembles of mixed states and representing the actual
quantity of quantum information produced by a source \cite{klasyka}. The problem whether the
passive information (equal to the number of needed qubits if the blind
coding schemes are considered) could be sometimes strictly greater than the
effective information associated with arbitrary coding schemes, remains open.
Finally we believe that the presented results will be useful  in further
investigations of the information content of ensemble of mixed states.

\begin{acknowledgements}
The author is grateful to R. Horodecki and P. Horodecki
for many helpful comments, discussions and technical remarks.
He also thanks T. Matsuoka and N. Watanabe for
interesting discussion on quantum information theory.
The financial support by Polish Committee for Scientific Research,
Contract No. 2 P03B 024 12 is gratefully acknowledged.
\end{acknowledgements}

\begin{appendix}
\section{}
Let $\Lambda: S(\hcal)\rightarrow S(\hcal)$ be a trace preserving completely
positive map, i.e. let it be of the following form \cite{Kraus}
\begin{equation}
\Lambda(\varrho)=\sum_iV_i \varrho V_i^\dagger.
\label{CP}
\end{equation}
Here $S(\hcal)$ is the set  of density matrices acting on the finite
dimensional Hilbert space $\hcal$, $V_i$'s are operators  satisfying
$\sum_i V_i^\dagger V_i=I$. It is known
that $\Lambda$ is of the form (\ref{CP}) if and only if it can be implemented
by means  of a unitary transformation over a larger system \cite{Kraus}
\begin{equation}
\Lambda(\varrho)={\rm Tr}_{\hcal'}U(\varrho\otimes P)U^\dagger.
\label{operacja}
\end{equation}
Here $P$ is pure state acting on the additional Hilbert space $\hcal'$;
$U$ is unitary transformation over the whole space $\hcal \otimes\hcal'$.
The form (\ref{operacja}) justifies the fact that the completely
positive trace preserving maps are identified with quantum operations.

Here we will proof the following proposition

{\bf Proposition.-} The distortion $D(\varrho,\sigma)$ does not increase under
quantum operations i.e. we have
\begin{equation}
D\left(\Lambda(\varrho),\Lambda(\sigma)\right)
\leq D(\varrho,\sigma).
\end{equation}

{\bf Proof.-} In view of the form (\ref{operacja}) it suffices to check whether
$D$ does not increase under the three components of the quantum operation:
unitary transformation, partial trace and the operation
$\varrho\rightarrow\varrho\otimes P$. As $D(\varrho,\sigma)=\|\varrho-\sigma\|$
depends only on the eigenvalues of the operator $A\equiv\varrho-\sigma$,
then it is unitarily invariant. Subsequently,  the operators 
$A$ and $ A\otimes P$ have
the same positive eigenvalues, so that
$D(\varrho\otimes P,\sigma\otimes P)=D(\varrho,\sigma)$.
Finally, suppose that $A$ acts on the Hilbert space $\hcal\otimes \hcal'$ and
has the spectral decomposition $A=\sum_i \lambda_i P_i$.
Let us estimate the trace norm of its partial trace
\begin{equation}
\|{\rm Tr}_{\hcal'}A\|=\|\sum_i\lambda_i\varrho_i\| \leq\sum_i|\lambda_i|\
\|\varrho_i\|=\sum_i|\lambda_i|=\|A\|.
\end{equation}
where $\varrho_i={\rm Tr}_{\hcal'} P_i$. Here we used triangle inequality
for the norm and the fact that $\|\varrho_i\|=1$. This completes the proof.
The proposition holds also in the case where the operation $\Lambda$  maps
$S(\hcal_1)$ into $S(\hcal_2)$ with different Hilbert spaces $\hcal_1$ and
$\hcal _2$.
\end{appendix}

\end{document}